\documentstyle[twoside,epsfig]{article}

\catcode`\@=11
\long\def\@makefntext#1{
\protect\noindent \hbox to 3.2pt {\hskip-.9pt
$^{{\eightrm\@thefnmark}}$\hfil}#1\hfill}		%CAN BE USED

\def\@makefnmark{\hbox to 0pt{$^{\@thefnmark}$\hss}}	%ORIGINAL
	
\def\ps@myheadings{\let\@mkboth\@gobbletwo
\def\@oddhead{\hbox{}
\rightmark\hfil\eightrm\thepage}
\def\@oddfoot{}\def\@evenhead{\eightrm\thepage\hfil
\leftmark\hbox{}}\def\@evenfoot{}
\def\sectionmark##1{}\def\subsectionmark##1{}}

\oddsidemargin=\evensidemargin
\newcounter{sectionc}\newcounter{subsectionc}\newcounter{subsubsectionc}
\renewcommand{\section}[1] {\vspace{12pt}\addtocounter{sectionc}{1}
\setcounter{subsectionc}{0}\setcounter{subsubsectionc}{0}\noindent
	{\tenbf\thesectionc. #1}\par\vspace{5pt}}
\renewcommand{\subsection}[1] {\vspace{12pt}\addtocounter{subsectionc}{1}
\setcounter{subsubsectionc}{0}\noindent
{\bf\thesectionc.\thesubsectionc. {\kern1pt \bfit #1}}\par\vspace{5pt}}
\renewcommand{\subsubsection}[1] {\vspace{12pt}\addtocounter{subsubsectionc}{1}
	\noindent{\tenrm\thesectionc.\thesubsectionc.\thesubsubsectionc.
	{\kern1pt \tenit #1}}\par\vspace{5pt}}
\newcommand{\nonumsection}[1] {\vspace{12pt}\noindent{\tenbf #1}
	\par\vspace{5pt}}

\newcounter{appendixc}
\newcounter{subappendixc}[appendixc]
\newcounter{subsubappendixc}[subappendixc]
\renewcommand{\thesubappendixc}{\Alph{appendixc}.\arabic{subappendixc}}
\renewcommand{\thesubsubappendixc}
	{\Alph{appendixc}.\arabic{subappendixc}.\arabic{subsubappendixc}}

\renewcommand{\appendix}[1] {\vspace{12pt}
        \refstepcounter{appendixc}
        \setcounter{figure}{0}
        \setcounter{table}{0}
        \setcounter{lemma}{0}
        \setcounter{theorem}{0}
        \setcounter{corollary}{0}
        \setcounter{definition}{0}
        \setcounter{equation}{0}
        \renewcommand{\thefigure}{\Alph{appendixc}.\arabic{figure}}
        \renewcommand{\thetable}{\Alph{appendixc}.\arabic{table}}
        \renewcommand{\theappendixc}{\Alph{appendixc}}
        \renewcommand{\thelemma}{\Alph{appendixc}.\arabic{lemma}}
        \renewcommand{\thetheorem}{\Alph{appendixc}.\arabic{theorem}}
        \renewcommand{\thedefinition}{\Alph{appendixc}.\arabic{definition}}
        \renewcommand{\thecorollary}{\Alph{appendixc}.\arabic{corollary}}
        \renewcommand{\theequation}{\Alph{appendixc}.\arabic{equation}}
        \noindent{\tenbf Appendix \theappendixc #1}\par\vspace{5pt}}
\newcommand{\subappendix}[1] {\vspace{12pt}
        \refstepcounter{subappendixc}
        \noindent{\bf Appendix \thesubappendixc. {\kern1pt \bfit #1}}
	\par\vspace{5pt}}
\newcommand{\subsubappendix}[1] {\vspace{12pt}
        \refstepcounter{subsubappendixc}
        \noindent{\rm Appendix \thesubsubappendixc. {\kern1pt \tenit #1}}
	\par\vspace{5pt}}

\newcommand{\textlineskip}{\baselineskip=13pt}
\newcommand{\smalllineskip}{\baselineskip=10pt}

\newcommand{\copyrightheading}[1]
	{\vspace*{-2.5cm}\smalllineskip{\flushleft
	{\footnotesize International Journal of Modern Physics D, #1}\\
	{\footnotesize \copyright\kern2pt World Scientific Publishing
	 Company}\\
	 }}

\newcommand{\publisher}[2]{{\begin{center}\footnotesize\smalllineskip
	Received #1\\
	Revised #2
	\end{center}
	}}

\def\abstracts#1#2#3{{
	\centering{\begin{minipage}{4.5in}\footnotesize\baselineskip=10pt
	\parindent=0pt #1\par
	\parindent=15pt #2\par
	\parindent=15pt #3
	\end{minipage}}\par}}

\newcommand{\bibit}{\nineit}

\renewenvironment{thebibliography}[1]
        {\frenchspacing
	 \ninerm\baselineskip=11pt
         \begin{list}{\arabic{enumi}.}
        {\usecounter{enumi}\setlength{\parsep}{0pt}
	 \setlength{\leftmargin 12.7pt}{\rightmargin 0pt}%FOR 1--9 ITEMS
         \setlength{\itemsep}{0pt} \settowidth
	{\labelwidth}{#1.}\sloppy}}{\end{list}}

\newcounter{itemlistc}
\newcounter{romanlistc}
\newcounter{alphlistc}
\newcounter{arabiclistc}
\newenvironment{itemlist}
    	{\setcounter{itemlistc}{0}
	 \begin{list}{$\bullet$}
	{\usecounter{itemlistc}
	 \setlength{\parsep}{0pt}
	 \setlength{\itemsep}{0pt}}}{\end{list}}

\newcommand{\fcaption}[1]{
        \refstepcounter{figure}
        \setbox\@tempboxa = \hbox{\footnotesize Fig.~\thefigure. #1}
        \ifdim \wd\@tempboxa > 5in
           {\begin{center}
        \parbox{5in}{\footnotesize\smalllineskip Fig.~\thefigure. #1}
            \end{center}}
        \else
             {\begin{center}
             {\footnotesize Fig.~\thefigure. #1}
              \end{center}}
        \fi}

\newcommand{\tcaption}[1]{
        \refstepcounter{table}
        \setbox\@tempboxa = \hbox{\footnotesize Table~\thetable. #1}
        \ifdim \wd\@tempboxa > 5in
           {\begin{center}
        \parbox{5in}{\footnotesize\smalllineskip Table~\thetable. #1}
            \end{center}}
        \else
             {\begin{center}
             {\footnotesize Table~\thetable. #1}
              \end{center}}
        \fi}

\def\@citex[#1]#2{\if@filesw\immediate\write\@auxout
	{\string\citation{#2}}\fi
\def\@citea{}\@cite{\@for\@citeb:=#2\do
	{\@citea\def\@citea{,}\@ifundefined
	{b@\@citeb}{{\bf ?}\@warning
	{Citation `\@citeb' on page \thepage \space undefined}}
	{\csname b@\@citeb\endcsname}}}{#1}}

\newif\if@cghi
\def\cite{\@cghitrue\@ifnextchar [{\@tempswatrue
	\@citex}{\@tempswafalse\@citex[]}}
\def\citelow{\@cghifalse\@ifnextchar [{\@tempswatrue
	\@citex}{\@tempswafalse\@citex[]}}
\def\@cite#1#2{{$\null^{#1}$\if@tempswa\typeout
	{IJCGA warning: optional citation argument
	ignored: `#2'} \fi}}

\def\pmb#1{\setbox0=\hbox{#1}
	\kern-.025em\copy0\kern-\wd0
	\kern.05em\copy0\kern-\wd0
	\kern-.025em\raise.0433em\box0}

\def\fnt#1#2{\footnotetext{\kern-.3em
	{$^{\mbox{\scriptsize #1}}$}{#2}}}

\def\fpage#1{\begingroup
\voffset=.3in
\thispagestyle{empty}\begin{table}[b]\centerline{\footnotesize #1}
	\end{table}\endgroup}

\def\runninghead#1#2{\pagestyle{myheadings}
\markboth{{\protect\footnotesize\it{\quad #1}}\hfill}
{\hfill{\protect\footnotesize\it{#2\quad}}}}
\font\tenrm=cmr10
\font\tenit=cmti10
\font\tenbf=cmbx10
\font\bfit=cmbxti10 at 10pt
\font\ninerm=cmr9
\font\nineit=cmti9

\font\eightrm=cmr8

\def\qed{\hbox{${\vcenter{\vbox{	          %HOLLOW SQUARE
   \hrule height 0.4pt\hbox{\vrule width 0.4pt height 6pt
   \kern5pt\vrule width 0.4pt}\hrule height 0.4pt}}}$}}

\begin{document}
\setlength{\textheight}{7.7truein}    %FOR 2ND PAGE ONWARDS

\runninghead{Initial Operation of the International Gravitational Event Collaboration}
%$\ldots$
{Initial Operation of the International Gravitational Event Collaboration}
%$\ldots$

\normalsize\textlineskip
\thispagestyle{empty}
\setcounter{page}{1}

\copyrightheading{}		%{Vol.~0, No.~0 (1999) 000--000}

\vspace*{0.88truein}

\fpage{1}
\centerline{\bf INITIAL OPERATION OF THE INTERNATIONAL }
\vspace*{0.035truein}
\centerline{\bf GRAVITATIONAL EVENT COLLABORATION}
%\footnote{For
%the title, try not to use more than 3 lines. Typeset the title
%in 10 pt Times Roman, uppercase and boldface.}}
\vspace*{0.37truein}
\centerline{\footnotesize G.A. PRODI,$^{a}$\footnote{presenting authors} \ I.S.
HENG,$^{b*}$\footnote{previously at: Department of Physics,
University of Western Australia, Nedlands, WA 6907} \ Z.A. ALLEN,$^b$ P. ASTONE,$^c$ L.
BAGGIO,$^d$ M. BASSAN,$^{e,f}$ }
\vspace*{0.015truein}
\centerline{\footnotesize D.G. BLAIR,$^g$ 
M. BONALDI,$^h$   P. BONIFAZI,$^{i,c}$  P. CARELLI,$^j$ M.
CERDONIO,$^d$  }
\vspace*{0.015truein}
\centerline{\footnotesize E. COCCIA,$^{e,f}$ L. CONTI,$^d$ C. COSMELLI,$^{k,c}$ V. CRIVELLI
VISCONTI,$^d$ S. D'ANTONIO,$^l$   }
\vspace*{0.015truein}
\centerline{\footnotesize V. FAFONE,$^l$ P. FALFERI,$^h$ P.
FORTINI,$^m$ S. FRASCA,$^{k,c}$ 
 W.O. HAMILTON,$^b$ }
\vspace*{0.015truein}
\centerline{\footnotesize E.N. IVANOV,$^g$ W.W. JOHNSON,$^b$ C.R. LOCKE,$^g$ A. MARINI,$^l$ 
V. MARTINUCCI,$^a$%\footnote{permanent address: }
}
\vspace*{0.015truein}
\centerline{\footnotesize E. MAUCELI,$^l$ 
M.P. McHUGH,$^b$ R. MEZZENA,$^a$ Y. MINENKOV,$^f$ I. MODENA,$^{e,f}$ 
}
\vspace*{0.015truein}
\centerline{\footnotesize G. MODESTINO,$^l$ A. MOLETI,$^{e,f}$ A. ORTOLAN,$^n$ G.V.
PALLOTTINO,$^{k,c}$ G. PIZZELLA,$^{e,l}$ 
}
\vspace*{0.015truein}
\centerline{\footnotesize E. ROCCO,$^d$ F. RONGA,$^l$ F. SALEMI,$^k$ G. SANTOSTASI,$^b$ L.
TAFFARELLO,$^o$ }
\vspace*{0.015truein}
\centerline{\footnotesize R. TERENZI,$^{i,f}$ M.E. TOBAR,$^g$ G. VEDOVATO,$^n$ A. VINANTE,$^a$ M. VISCO,$^{i,f}$ 
S. VITALE,$^a$ }
\vspace*{0.015truein}
\centerline{\footnotesize L. VOTANO,$^l$ J.P. ZENDRI$^o$ }
\vspace*{0.015truein}
\centerline{\footnotesize\it $^a$ Dipartimento di Fisica, Universit\`a
di Trento, and I.N.F.N., Gruppo Collegato di Trento,}
\baselineskip=10pt
\centerline{\footnotesize\it I-38050 Povo, Trento, Italy}
\baselineskip=10pt
\centerline{\footnotesize\it $^b$ Department of Physics and Astronomy, 
Louisiana State University,}
\baselineskip=10pt
\centerline{\footnotesize\it Baton Rouge, Louisiana 70803}
\baselineskip=10pt
\centerline{\footnotesize\it $^c$ I.N.F.N., Sezione di Roma1, P.le A.Moro 2, }
\baselineskip=10pt
\centerline{\footnotesize\it I-00185, Roma, Italy}
\baselineskip=10pt
\centerline{\footnotesize\it $^d$ Dipartimento di Fisica, Universit\`a
di Padova, and I.N.F.N., Sezione di Padova,}
\baselineskip=10pt
\centerline{\footnotesize\it Via Marzolo 8, 35131 Padova, Italy}
\baselineskip=10pt
\centerline{\footnotesize\it $^e$ Dipartimento di Fisica, Universit\`a 
di ``Tor Vergata'', Via O.Raimondo,}
\baselineskip=10pt
\centerline{\footnotesize\it  I-00173 Roma, Italy}
\baselineskip=10pt
\centerline{\footnotesize\it $^f$ I.N.F.N. Sezione di Roma2, Via O.Raimondo,}
\baselineskip=10pt
\centerline{\footnotesize\it  I-00173 Roma,Italy}
\baselineskip=10pt
\centerline{\footnotesize\it $^g$ Department of Physics,
University of Western Australia,}
\baselineskip=10pt
\centerline{\footnotesize\it Nedlands, WA 6907
Australia}
\baselineskip=10pt
\centerline{\footnotesize\it $^h$ Centro di Fisica degli Stati Aggregati, I.T.C.- C.N.R., and I.N.F.N., Trento, }
\baselineskip=10pt
\centerline{\footnotesize\it I-38050 Povo, Trento, Italy}
\baselineskip=10pt
\centerline{\footnotesize\it $^i$ Istituto Fisica Spazio Interplanetario, 
C.N.R., Via Fosso del Cavaliere,}
\baselineskip=10pt
\centerline{\footnotesize\it I-00133 Roma, Italy}
\baselineskip=10pt
\centerline{\footnotesize\it $^j$ Dipartimento di Fisica, Universit\`a de L'Aquila, and I.N.F.N., }
\baselineskip=10pt
\centerline{\footnotesize\it L'Aquila,Italy}
\baselineskip=10pt
\centerline{\footnotesize\it $^k$ Dipartimento di Fisica, Universit\`a 
``La Sapienza'', P.le A.Moro 2,}
\baselineskip=10pt
\centerline{\footnotesize\it  I-00185, Roma, Italy}
\baselineskip=10pt
\centerline{\footnotesize\it $^l$ Laboratori Nazionali di Frascati,
Istituto Nazionale di Fisica Nucleare,
Via E.Fermi 40, }
\baselineskip=10pt
\centerline{\footnotesize\it I-00044, Frascati, Italy}
\baselineskip=10pt
\centerline{\footnotesize\it $^m$ Dipartimento di Fisica, Universit\`a di Ferrara, and I.N.F.N., Sezione di Ferrara}
\baselineskip=10pt
\centerline{\footnotesize\it  I-44100 Ferrara, Italy}
\baselineskip=10pt
\centerline{\footnotesize\it $^n$ Laboratori Nazionali di Legnaro,
Istituto Nazionale di Fisica Nuclare,}
\baselineskip=10pt
\centerline{\footnotesize\it 35020 Legnaro, Padova, Italy}
\baselineskip=10pt
\centerline{\footnotesize\it $^o$ Sezione di Padova, Istituto Nazionale di Fisica Nucleare, via Marzolo 8,}
\baselineskip=10pt
\centerline{\footnotesize\it  I-35131 Padova, Italy}
\vspace*{0.225truein}
\publisher{(received date)}{(revised date)}

\vspace*{0.21truein}

\abstracts{
The International Gravitational Event Collaboration, IGEC, is a
coordinated effort by research groups operating gravitational wave
detectors
working towards the detection of millisecond bursts of gravitational waves. Here we
report on the current IGEC resonant bar observatory, its data analysis
procedures, the main properties of the first exchanged data set. Even
though the available data set is not complete, in the years 1997 and
1998 up to four detectors were operating simultaneously. Preliminary results are mentioned. 
}{}{}

%\vspace*{10pt}
%\keywords{The contents of the keywords}

%\textlineskip			%) USE THIS MEASUREMENT WHEN THERE IS
%\vspace*{12pt}			%) NO SECTION HEADING

\vspace*{1pt}\textlineskip	%) USE THIS MEASUREMENT WHEN THERE IS
\section{The International Gravitational Event Collaboration}	%) A SECTION HEADING
\vspace*{-0.5pt}
\noindent
One of the most relevant scientific objectives for resonant detectors of gravitational
waves (gw) is the search for short gw bursts, emitted during
the gravitational collapse of stars or the final evolution of 
coalescing binaries.\cite{sources} To ensure the confidence of a detection,
it is necessary to compare the observations made by multiple
detectors with uncorrelated noise.
This has already been done in
the past years among pairs of cryogenic detectors with common time periods of
observation of about a
semester.\cite{89-lsuromastanford}$^,$\cite{99-allegroexplorer} A few
days of observation have been
reported also for three simultaneously operating 
detectors.\cite{89-lsuromastanford}
In these attempts, the gw search consisted of a time
coincidence analysis among the candidate signals reported by the
different detectors and no statistically significant excess of
coincidences was found.

An increase of the number of cryogenic resonant detectors in simultaneous 
operation in recent years has greatly improved the chance of making a 
confident detection of gw bursts. In fact,
the International
Gravitational Event Collaboration, IGEC, currently consists of the research
groups operating the five cryogenic bar detectors 
ALLEGRO,\cite{AL} AURIGA,\cite{AU} EXPLORER,\cite{EX} NAUTILUS\cite{NA} and
NIOBE.\cite{NI} The IGEC was established in July 1997 
with an agreement\cite{IGEC} for
setting up a common search for gravitational wave bursts of
duration of the order of $1 ms$. This agreement sets the guidelines for
the data exchange procedure among the participating groups and the IGEC
scientific policy, whose most relevant aspects are:
\begin{itemlist}
\item each  group has responsibility to make available to IGEC its list of
candidate gravitational wave
events,
\item a unanimous agreement of the member groups is
required to make public the results based on the IGEC data exchange,
\item IGEC is open to new data taking research groups.
\end{itemlist}

%The first implementation of the agreement was carried out in the
%proceeding year. 
\noindent
In 1999 the
first IGEC analysis of the 1997-1998 data was performed and some initial results
will be presented in the following
sections. At the time of this analysis, not all the
1997-1998 data had been exchanged. Despite the incomplete data set, the
simultaneous operation of four gravitational wave detectors was achieved
for the first time.

\subsection{Data exchange protocol}
\noindent
The IGEC data exchange
procedure is  aimed at searching for coincident
excitations at different detectors. For each detector, a list of
candidate events, each describing a $\delta$-like gravitational wave
 excitation of the detector, is provided by the corresponding
 research group. The IGEC protocol
requires that in each list the 
 candidate event rate be at most of the order of
$100/day$, to limit the expected rate of accidental coincidences.
Currently, the research groups have been exchanging event lists  relative to
the past three years, but a
future goal is to implement an automatic exchange day by day.

The event lists are then made available to the IGEC collaboration as
files under a common protocol,\cite{IGEC} open for future extensions.
Under this protocol, it is mandatory for each
detector to provide the minimum set of information needed to describe a
$\delta$-like gw excitation of the detector for each event; namely its
Universal Time of arrival, the Fourier component $H_o$ of its amplitude
in $Hz^{-1}$ and the detector noise level at that time. Another
mandatory requirement is the declaration of the effective observation
time of each detector, so that the IGEC observation time can be
calculated. Optional information, such as the time of threshold crossing of the
detector output, the duration of the event and its statistical compliance
to a $\delta$-like gw excitation, can also be exchanged.

\subsection{IGEC gravitational wave observatory}
\noindent
The relevant parameters of the five resonant bar detectors of
the IGEC observatory in the years 1997-1998 are summarised in Table 1.
The detectors are sensitive to gw signals  in a typical bandwidth
 of the order of $1 Hz$ around each one of the two 
resonances of the detector, which
are close to $700 Hz$ for NIOBE and close to $900 Hz$ for all other
detectors. The relationship between the Fourier amplitude $H_0$,
averaged on the two resonant frequencies of the detector, and 
the energy $E_s$ deposited by the g.w. burst on the bar, is given by:
\begin{equation}
H_0 = \frac{1}{4 L_{bar} {\nu_0}^2} \sqrt{E_s/ M_{bar}}     
\end{equation}
where $L_{bar}$ is the bar length, $M_{bar}$ its mass, $\nu_0$ the mean
of the detector resonance frequencies.

%\vspace*{4pt}   %only when needed
\begin{table}[htbp] 
\tcaption{Main characteristics of the five resonant bar detectors in
1997-1998. The reported misalignment is the angle between the bar axis
and an optimal direction whose overall misalignment from the detectors
is minimal. $Q_{\pm}$ is the typical quality factor of the resonances}
%\centerline{\footnotesize NP}
\centerline{\footnotesize\smalllineskip
\begin{tabular}{l c c c c c c}\\
\hline
{ANTENNA} &{ALLEGRO} &{AURIGA} &{EXPLORER} &{NAUTILUS} &{NIOBE}\\
\hline
{Material} & $AL 5056$ & $AL 5056$ &$AL 5056$ &$AL 5056$ &$Nb$\\
{Mass [$kg$]} & 2296 & 2230 & 2270 & 2260 & 1500\\
{Length [$m$]} & 3.0 & 2.9 & 3.0 & 3.0 & 2.75\\
{Mode - [$Hz$]} &895 &912 &905 &908 &694\\
{Mode + [$Hz$]} &920 &930 &921 &924 &713\\
{$Q_{\pm}$ [$10^6$]} &2 &3 &1.5 &0.5 &20\\
{Temp. [$K$]} &4.2 &0.2 &2.6 &0.1 &5.0\\
{Longitude} & $268^\circ 50' E$ & $11^\circ 56' 54" E$ & $6^\circ 12' E$ & $12^\circ 40' 21" E$ & $115^\circ 49' E$ \\
{Latitude} & $30^\circ 27' N$ & $45^\circ 21' 12" N$ & $46^\circ 27' N$ & $41^\circ 49' 26" N$ & $31^\circ 56' S$ \\
{Azimuth} & $40^\circ W$ & $44^\circ E$ & $39^\circ E$ & $44^\circ E$ & $0^\circ$ \\
{Misalignment [$deg$]} & 9 & 4 & 2 & 3 & 29\\
\hline\\
\end{tabular}}
\end{table}
%misalignment rispetto al quello che minimizza disallineamento rispetto
%tutte tranne Niobe 
%auriga 		45° 21' 12" N, 	11° 56' 54" E, azimuth dal N 44°
%nautilus	41° 49' 26" N, 	12° 40' 21" E, 		44°
%explorer	46° 27' N, 	6° 12' E,		39°
%allegro		30° 27' N, 	268° 50' E,		-40°
%niobe		31° 56' S, 	115° 49' E,		0°
The typical  thresholds used for selecting the events in
 the 1997-1998 burst search
have been in the range $H_0 \simeq 2 \-- 6 \times 10^{-21} Hz^{-1}$.
The corresponding strain amplitude of the gw can be computed assuming
a model for the burst shape: for the conventional $\sim 1 ms$ burst, the
Fourier component $H_0$ should be multiplied by $10^3 Hz$ to get the maximum
strain amplitude $h$.

To maximise the chances of a coincidence detection, the bars have been
oriented to be approximately parallel to one another. Neglecting the
polarisation effects, the gw amplitude at the detector is $H_0 = H_{gw}
sin^2\theta(t)$, where $H_{gw}$ is the incident gw amplitude and
$\theta(t)$ is the angle between the bar axis and the direction of the
source. In this configuration of the observatory, the relative
misalignments reported in Tab. 1 for ALLEGRO, AURIGA, EXPLORER and
NAUTILUS disperse their $sin^2\theta(t)$ responses by at most a few \%,
while for NIOBE the dispersion is up to a few tenths. Figure 1 shows as
an example the resulting amplitude efficiency for the observation of the
Galactic Center as the Earth rotates. Since their values of
$sin^2\theta(t)$ are simultaneously above 0.7 for about 60\% of the
time, we point out that this configuration of the observatory ensures a
rather good and coherent coverage of the central galactic mass during
time.

\begin{figure}[htbp] %ORIGINAL SIZE: width=1.4TRUEIN; height=1.5TRUEIN
\vspace*{13pt}
\centerline{\psfig{file=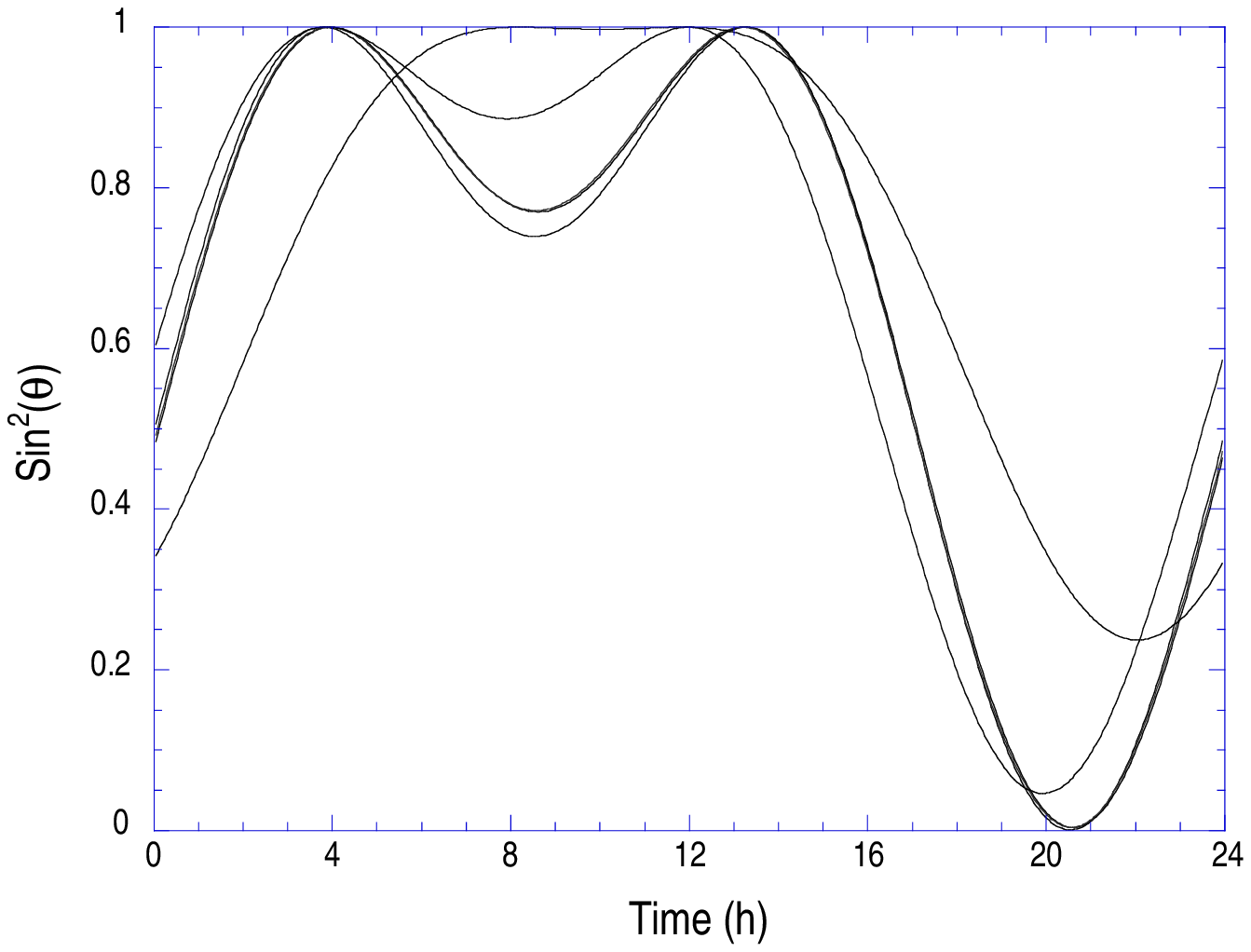,width=7cm}} %100 percent
\vspace*{13pt}
\fcaption{Amplitude response $sin^2\theta(t)$ of the five IGEC detectors
 versus Universal Time for signals incoming from the Galactic Center during
one day.}
\end{figure}

\section{The Data Set} 
\noindent 
Each IGEC group
independently developed a data acquisition and an optimum  filtering
procedure for a $\delta$-like
gw excitation. These procedures differ greatly in the
methods, in particular some of them filter for the energy released in
the bar (ALLEGRO,\cite{AL} NIOBE\cite{NI}) while the others filter for
the amplitude and phase of the strain excitation of the bar (AURIGA,\cite{AURanalisi}
EXPLORER and NAUTILUS\cite{ROGanalisi}). The output correlation
time of the filters ranges from a few tenths  up to a
few seconds. 

A search for maxima on the filter outputs is then used to
identify the time and amplitude of the candidate events, which are
exchanged only if their amplitude exceeds a selected threshold relative to
the noise level. For the current detectors, these thresholds span in
the range of signal-to-noise ratio
$SNR\simeq 3 \-- 5$ in amplitude. Of the events overcoming the
threshold, some are rejected as spurious by different methods. 
AURIGA implements  a $\chi^2$ test on
the compliance of the single detected excitations with the expected
template of a gw burst.\cite{chi2} In fact, the filtering procedure implemented
is equivalent to a maximum likelihood fit of a signal model to the data and
the goodness of the fit is statistically tested for each candidate event; those
events not passing the test are rejected.
All the other detectors implement spurious signals
rejection using  sensors of ambient disturbances. In addition, EXPLORER
and NAUTILUS reject the events when the local noise is above
a certain threshold.

The effective observation times for each detector has been defined by
vetoing, {\it a priori}, time periods of detector maintenance or
malfunctions and times when the detector was excited by the local
environment as determined by the local experimentalists. After
filtering, AURIGA implements a second level of vetoes {\it a posteriori}
to reject further periods of unsatisfactory performance due to a lack of
self consistency of its data analysis, that is when its noise fails to
be compliant with the modeled one used to build its filtering
procedure.\cite{analisiAU}
   
\begin{table}[htbp]
\tcaption{Left: total observation time $T_{obs}$ and rate of events
$R_{evt}$ for each detector. Right:  net common observation time $T_{N}$
 when at least $N_{detectors}$ were simultaneously operating.}
\centerline{\footnotesize\smalllineskip
\begin{tabular}{l c c @{\hspace{10mm}} || @{\hspace{10mm}} r c}\\
\hline
\emph{Detector} &
\emph{$T_{obs}~ (day)$} &
\emph{$R_{evt} ~(events/day)$} &
\emph{$N_{detectors}$} &
\emph{$T_{N}~ (day)$}\\
\hline
{ALLEGRO}  &405.7 &112.9 &  1 &  625.0\\
{AURIGA}   &153.0 &175.3 &  2 &  260.4\\
{EXPLORER} &137.5 &150.7 &  3 &  \phantom089.7\\
{NAUTILUS} &108.5 &80.8  &  4 &  \phantom015.5\\
{NIOBE}    &185.9 &14.0  &  5 &  \phantom0\phantom00\\
\hline\\
\end{tabular}}
\end{table}
The amplitude distributions of all the 1997-1998 exchanged events for each
detector are shown in Fig.2. The effective observation time of the exchanged
data up to now is summarised in Table 2, together with the mean rate of
exchanged events. The net observation time with at least four, three and
two detectors simultaneously operating has been respectively 15.5, 90
and 260 days. We expect that the three-way observation time will
increase significantly as the exchanged data set will become complete.
The ALLEGRO detector has been showing the best duty cycle, close to
$100 \%$ on the exchanged data period, as well as the most stationary
noise performance with respect to the other detectors. 

\begin{figure}[htbp] %ORIGINAL SIZE: width=1.4TRUEIN; height=1.5TRUEIN
\vspace*{13pt}
\centerline{\psfig{file=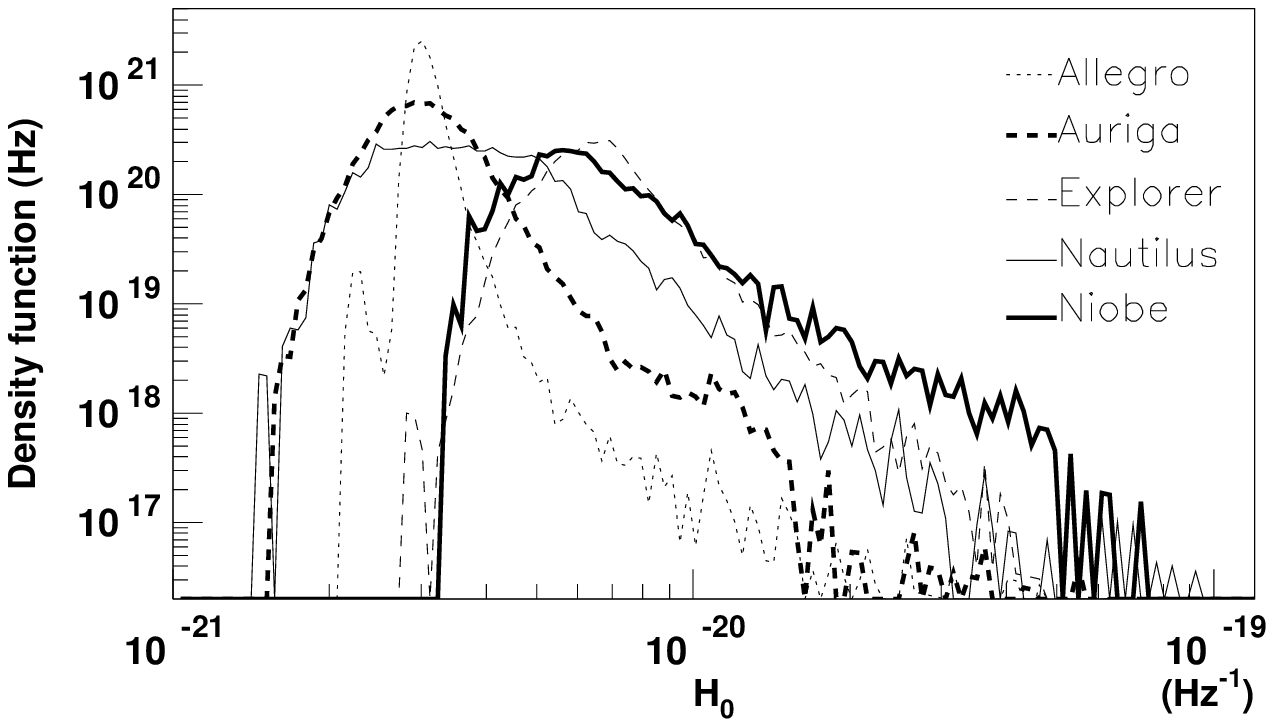,width=12cm}} %100 percent
\vspace*{13pt}
\fcaption{Density functions of the amplitude of all the exchanged events in 1997-1998
for each detector. The densities are normalised to unit area and are estimated 
from the amplitude histograms of the events.
}
\end{figure}

\section{Analysis of Time Coincidences} 
\noindent 
A search for two,
three and four-fold coincidences was carried out on the exchanged data.
In the analysis reported here a $M$-fold coincidence is observed if the
estimated arrival times $t_i$ at the $M$ detectors are all $|t_i-t_j|\leq1s$.
This figure has been chosen as a compromise between the demands for a
small accidental background and for a low false dismissal probability.\footnote{ 
a new approach for multiple time coincidence analysis has
been proposed by the Rome group.\cite{roberto}}\ 
In fact, the measured uncertainties $t_w $ on the estimated
arrival times of a burst at each detector are quite similar and are
selected to be $ \pm 0.5 s$. This corresponds to a maximum false
dismissal of a few \% for the exchanged events, even at low $SNR$. The
maximum separation among coincident events is therefore set to $2 t_w =
1 s$.

A preliminary search for three and four-fold coincidences shows none. A
detailed analysis is in progress.
The two-fold coincidences found for each pair of detectors are shown in
Table 3: in all the cases they correspond to the estimated accidental background.

\begin{table}[htbp]
\tcaption{Preliminary results of two way coincidence analysis for the 1997-1998
data. The abbreviations, AL, AU, EX, NA and NI stand for
ALLEGRO, AURIGA, EXPLORER NAUTILUS and NIOBE respectively. $n_{c}$ is
the found number of coincidences, $<n_{a}>$ and $<n_{a}>_{theory}$ the expected accidental ones
respectively from the time shift
method and Eq.\ref{eq:one}, $P(n\ge n_{c}|<n_{a}>)$ the calculated
probability for the coincidences to be $\ge n_{c}$, and $T_{obs}$ the common observation time
of the pairs of detectors.
}
\centerline{\footnotesize\smalllineskip
\begin{tabular}{c c c c c c}\\
\hline
\emph{Detectors} &
\emph{$n_{c}$} &
\emph{$<n_{a}>$} &
\emph{$<n_{a}>_{theory}$} &
\emph{$P(n\ge n_{c}|<n_{a}>)$} &
\emph{$T_{obs}$}\\
\hline
{AL-AU} &42 &46.6 &45.1 &0.77 &103.8\\
{AL-EX} &27 &31.2 &30.9 &0.80 &100.7\\
{AL-NA} &17 &21.6 &21.0 &0.87 &98.9\\
{AL-NI} &1  &0.9  &1.0  &0.61 &27.1\\
{AU-EX} &14 &20.3 &19.2 &0.94 &44.1\\
{AU-NA} &4  &4.2  &4.2  &0.60 &18.3\\
{AU-NI} &1  &2.3  &2.1  &0.90 &37.0\\
{EX-NA} &5  &7.0  &5.7  &0.83 &37.5\\
{EX-NI} &1  &1.1  &1.0  &0.65 &18.9\\
\hline\\
\end{tabular}}
\end{table}

Two methods for estimating the rate of accidental coincidences have been
applied for the pairs of detectors: i) performing several time shifts of events times of one
detector with respect to the other and then looking for
coincidences\cite{89-lsuromastanford} (the resulting accidental
coincidences are reported as $<n_a>$ in Tab.3); ii) assuming
stationary Poisson distributions of event times and using the mean
measured rates of events for each detector during the common observation
time. The latter method predicts a number of accidental $M$-fold
coincidences given by\cite{accidental}
\begin{equation} 
<n_{a}>_{theory} =
M\frac{(2t_{w})^{M-1}}{T^{M-1}_{obs}}\prod^{M}_{i=1}n_{i}
,
\label{eq:one} 
\end{equation} 
where $M$ is the number of detectors,
$T_{obs}$ their common observation time, $t_{w}$ the time window
describing the timing accuracy of each detector, $n_{i}$ the number of
exchanged candidate events of the $i^{th}$ detector during $T_{obs}$. The
agreement of both estimates of the accidental background of coincidences
is well within the statistical uncertainties for the detector pairs. In
addition, the observed coincidences, $n_c$, correspond to both estimates
of the accidental coincidence background. This implies that no excess
coincidences were observed.

Using Eq.\ref{eq:one}, we also performed a preliminary analysis of the
rates of accidental coincidences for three and four-fold configurations
of the IGEC observatory. The most relevant result here is that the
statistical significance of three-fold and four-fold time concidences among
the current IGEC detectors improves by order of magnitudes.

\section{Future Plans and Conclusions}
\noindent
This first IGEC joint analysis has shown that, with the current detector
performances and the selected coincidence time window, 
at least three detectors simultaneously operating can perform 
an autonomous search for gw bursts
with a very low false alarm rate even at signal-to-noise ratio as low as
$3 \-- 5$ in single detectors.
Therefore, the continuation of the IGEC international effort is very
strongly motivated. Moreover, the joint work of the participating
groups has instigated
efforts to co-ordinate data analysis techniques between the different
groups.

No coincidence above the expected 
accidental background were found in
this preliminary analysis. Work is in progress to complete the analysis to 
three and four-fold coincidences, as well as  to set upper limits on
the rate of incoming gw bursts and on the amplitude of single gw bursts
associated with selected time windows of astrophysical interest. 

Improvements in the detector performances will lead to increase the
sensitivity of the IGEC observatory in two respects. On one side the
thresholds for gw burst search will be lowered without increasing the
level of the accidental background rate. On the other side the effective
bandwidths of the detectors will be also widened, thus decreasing the
uncertainty in the estimated arrival time of a gw burst. The latter will
allow the lowering of the rate of
accidental coincidences and, above all, it will give the opportunity to measure the
propagation speed and direction of the incoming gw.

In this framework, we think that the participation within the IGEC
observatory of the interferometric detectors as they will begin
observations would constitute a very important stage towards the
establishment of the future worldwide observatory for gravitational
waves.

\nonumsection{Acknowledgements}
\noindent
The ALLEGRO group was supported by the National Science Foundation and
LSO, the NIOBE group by the Australian
Research Council. The Italian groups were supported in part by a grant from MURST-COFIN'97.
The Rome group thanks 
F.Campolungo, G.Federici, R.Lenci, G.Martinelli, E.Serrani, R.Simonetti
and F.Tabacchioni for their precious technical assistance.

\nonumsection{References}

\end{document}